\documentclass[a4paper,12pt]{article}

\ifx\pdfoutput\undefined%
   \IfFileExists{graphicx.sty}{\RequirePackage[dvips]{graphicx}
   \DeclareGraphicsExtensions{.eps,.ps}}{}
\else%
   \ifnum\pdfoutput=0
      \IfFileExists{graphicx.sty}{\RequirePackage[dvips]{graphicx}
      \DeclareGraphicsExtensions{.eps,.ps}}{}
   \else
      \IfFileExists{graphicx.sty}{\RequirePackage[pdftex]{graphicx}
      \DeclareGraphicsExtensions{.pdf,.png,.jpg}}{}
   \fi
\fi 

\newcommand{\IJNE}{{\it Int. J. Numer. Meth. Eng.} }

\newcommand{\MT}{{\it Metall. Trans. } }

\newcommand{\IJHPCA}{{\it International Journal of High Performance Computing Applications} }

\newcommand{\IJCMR}{{\it Int. Journal of cast metals Research} }
\newcommand{\ETNA}{{\it Electronic Transaction on Numerical Analysis} }
\newcommand{\MSMSE }{{\it Modelling Simul. Mater. Sci. Eng.} }
\newcommand{\JPDC }{{\it Journal of Parallel and Distributed Computing } }

\usepackage{harvard}

\begin{document}


\begin{center}

\textsc{Improvement Cache Efficiency of Explicit Finite Element
Procedure and its Application to Parallel Casting Solidification
Simulation}

\textsc{R. Tavakoli $^{\dag}$}

\begin{small}

$^\dag$ Material Science and Engineering Department, Sharif
University of Technology, P.O. Box 11365-9466, Tehran, Iran,
 email: tav@mehr.sharif.edu, rohtav@gmail.com\\

April 2006\\

\end{small}

\end{center}

\vspace{3mm}




\vspace*{2mm} {\Large\bf\noindent Abstract\\}

\noindent{A simple method for improving cache efficiency of serial
and parallel explicit finite procedure with application to casting
solidification simulation over three-dimensional complex geometries
is presented. The method is based on division of the global data to
smaller blocks and treating each block independently from others at
each time step. A novel parallel finite element algorithm for
non-overlapped element-base decomposed domain is presented for
implementation of serial and parallel version of the presented
method. Effect of mesh reordering on the efficiency is also
investigated. A simple algorithm is presented for high quality
decomposition of decoupled global mesh. Our result shows 10-20 \%
performance improvement by mesh reordering and 1.2-2.2 speedup with
application of the presented cache efficient algorithm (for serial
and parallel versions). Also the presented parallel solver (without
cache-efficient feature) shows nearly linear speedup on the
traditional Ethernet networked Linux cluster.}



\section{Introduction}

{\large\bf\noindent Finite element analysis\\} \noindent Many
physical phenomena such as transient heat transfer and fluid flow
are governed by parabolic partial differential equations (PDEs).
Finite element method (FEM) (Zienkiewicz and Taylor, 2000) is one of
the powerful numerical techniques for solution of PDEs. In FEM the
spatial domain is discretized with arbitrary unstructured grid, and
the temporal domain is usually discretized with explicit or implicit
finite difference method. The explicit time integration method has
certain advantages in contrast with implicit method, such as:
simplicity of implementation, smaller memory requirement and better
parallelism and scalability. However, they suffer the severely
restricted time step size from stability requirement. In implicit
time integration method, there is not any time step restriction. But
in each time step one has to solve a global linear or non-linear
algebraic system of equations that is severely memory and time
consuming procedure and difficult to implementation specially for
distributed memory architecture.

Selection between explicit and implicit time integration methods
from efficiency viewpoint is not easy task and is function of
several parameters such as: behavior of PDEs, size of problem and
available computational resource. Generally it can be said that for
highly non-linear problems the explicit method is preferable because
usually large time increment leads to increasing overall
computational cost, due to increasing number of non-linear
iterations. Also, in some problems there are additional time step
limitation criteria that don't permit application of the large time
increment. For example application of large time step leads to
interface tracking inaccuracy or instability in moving boundaries
and impact problems and missing latent heat effect (and consequently
energy unbalance) in the phase change problems.

\vspace{10mm} {\large\bf\noindent Cache efficient algorithm:
Background\\} \noindent Years ago, knowing how many floating-point
operations (FLOPS) were used in a given algorithm (or code) provided
a good measure of the running time of a program. This was a good
measure for comparing different algorithms to solve the same
problem. This is no longer true (Douglas et al., 2000a).

The growing speed gap between processor and memory has led to the
development of memory hierarchies and to the widespread use of
caches in modern processors (Sellappa and Chatterjee, 2004). The
highest level is the most expensive, smallest and fastest of all
layers. Lower levels are progressively larger, cheaper and slower.
The layers (from highest to lowest) are registers, cache and main
memory. The cache is itself a hierarchy L1, L2, … (levels of cache).
One or two levels of cache is typical. The L1 cache is smaller and 2
to 6 times faster than the L2 cache. The L2 cache in turn is smaller
than but still 10 to 40 times faster than main memory. The data held
in any level of cache is usually a subset of the next larger level.
The smallest block of data moved in and out of a cache is a cache
line. A cache line holds data, which is stored contiguously in main
memory. If the data that the CPU requests are in a cache line, it is
called a cache hit. Otherwise data must be copied from a lower level
of memory into a line, a cache miss. The hit rate is the ratio of
cache hits to total CPU requests. The miss rate is 1 minus the hit
rate. It is clearly advantageous to maximize the number of cache
hits, since cache is faster than main memory in fulfilling the CPU's
request (Hu, 2000).

Therefore the speed of a code depends increasingly on how well the
cache structure is exploited and the number of cache misses provides
a better measure for comparing algorithms than the number of FLOPS.
Unfortunately, estimating cache misses is difficult to model a
priori and only somewhat easier to do a posteriori (Douglas et al.,
2000a). When a program references a memory location, the data in the
referenced location and nearby locations are brought into a cache.
Any additional references to data before these are evicted from the
cache will be one or two orders of magnitude faster than references
to main memory. So increasing data locality leads to decreasing
cache miss. A program is said to have good data locality if, most of
the time, the computer finds the data referenced by the program in
its cache. The locality of a program can be improved by changing the
order of computation and/or the assignment of data to memory
locations so that references to the same or nearby locations occur
near to each other in time during the execution of the program (
Strout et al., 2004).

When we deal with finite element analysis (FEA) on unstructured
mesh, the data connectivity has unstructured sparse structure and
data storage must involve indirect addressing. So the cache miss is
considerably increased due to the irregularity of many of the memory
accesses. Many researchers have explored the benefits of reordering
sparse matrices resulted from FEA (Pinar and Heath,1999; Im and
Yelick, 1999; Oliker et al., 2000; Lohner, 2001,). In this method
the bandwidth of global matrix (resulted from FEA) is reduced with
node renumbering which decreases chance of cache miss during
computation. But this method shows little benefit for large-scale
problems on uniprocessors (Im and Yelick, 1999; see also
experimental results section).

Related works in the category of cache efficient numerical algorithm
were focused to improvement cache efficiency of solution of system
of linear equations with stationary iterative methods (e.g.
Gauss-Seidel as direct solver or as a smoother in the multigrid
methods). So related works fall in the category of cache efficient
implicit FEA.

Work related to performance transformations for stationary iterative
methods can be categorized by whether it deals with regular or
irregular meshes and whether it attempts to improve intra-iteration
locality and/or inter-iteration locality. Another important
distinction is between code transformations that have been automated
in a compiler versus programming transformations that require some
domain-specific knowledge and are currently applied by hand.

Tiling is the process of decomposing a computation into smaller
blocks and doing all of the computing in each block one at a time.
Tiling is an attractive method for improving data locality. In some
cases, compilers can do this automatically. However, this is rarely
the case for realistic scientific codes. In fact, even for simple
examples, manual help from the programmers is, unfortunately,
necessary (Douglas et al., 2000b).

Traditional tiling (Wolfe, 1987; Irigoin and Triolet, 1988; Gannon
et al., 1988; Wolfe, 1989; Wolf and Lam, 1991; Carr and Kennedy,
1992; McKinley et al., 1996) can be applied to a perfect loop nest
that traverses the unknowns associated with a regular mesh, provided
the memory references and loop boundaries are affine and the
unknowns are ordered in a way that allows a compile-time analysis to
find a legal tiling. In particular, after the enabling
transformation skewing is applied, tiling is often applicable to
Gauss-Seidel and SOR over a regular mesh. For Jacobi over a regular
mesh, tiling transformations developed for imperfectly nested loops
can be used (Song and Li, 1999; Ahmed et al., 2000). Another issue
involved in tiling computations on regular meshes is how to
determine the tiling and array padding parameters (Rivera and Tseng,
2000). If other compiler transformations, such as skewing, function
inlining and converting 'while' loops to 'for' loops, are used, then
it is possible to apply tiling transformations for imperfectly
nested loops to stationary iterative methods to achieve
inter-iteration locality on regular meshes.

Increasing inter-iteration locality through programming
transformations for iterative stationary methods on structured
meshes is explored by Leiserson et al. (1997) with blocking covers
technique, Stals and Rüde (1997) with cache blocking along one
dimension for two-dimensional problem, Kodukula et al. (1997) with
data-centric multi-level blocking technique, Bassetti et al. (1998)
with stencil optimization techniques, Bassetti et al. (1999) with
combining blocking covers with stencil optimization techniques,
Povitsky (1999) with Wavefront cache-friendly algorithm, Pugh and
Rosser (1999) with iteration space slicing, Wonnacott (2002) with
time skewing, Douglas et al. (2000a) and Sellappa and Chatterjee
(2004) with two-dimensional cache blocking techniques. Such
programming transformations are explored extensively due to the
prevalence of iterative regular mesh computations that compilers do
not tile because the necessary combinations of enabling
transformations are not done automatically. For unstructured grids,
Cache blocking (Douglas et al., 2000$^a$) and sparse tiling(Strout
et al., 2004) are two such run-time reordering transformations
developed for the stationary iterative methods. Im (2000), Im and
Yelick (2001) and Im et al. (2004) have developed a code generator,
SPARSITY, that improves the intra-iteration locality for the general
sparse matrix operations by register blocking and cache blocking
techniques.

There has also been work on compiler-generated inspectors/executors
for improving intra-iteration locality of irregular problems (Ding
and Kennedy, 1999; Mitchell et al., 1999; Han and Tseng, 2000;
Mellor-Crummey et al., 2001). These papers describe how a compiler
can analyze non-affine array references in a loop and generate the
inspectors and executors for performing run-time data and iteration
reordering. These transformations can be applied to the inner loops
of Jacobi implemented for sparse matrix formats, but not to
Gauss-Seidel or SOR due to the intra-iteration data dependences.

\vspace{10mm} {\large\bf\noindent Outline of present work\\}
\noindent As stated in the above section most of works in the
category of cache efficient algorithms have been attended to
implicit and iterative methods. So extension of those ideas to
explicit and non-iterative methods (for developing cache efficient
algorithms) can be valuable work, which is a part of the present
study.

The scope of the present study is high performance implementation of
explicit finite element procedure for solution of large-scale
problems, which are governed with non-linear parabolic PDEs. In this
study our focus is on the simulation casting solidification, which
is a non-linear phase change problem. For this purpose at first step
we try to increase cache efficiency of explicit finite element
procedure for single processor execution and then extend it to
distributed memory parallel execution. Although we focus on the
special problem but the introduced idea is general and extendable to
other applications.

The original contributions of this paper are: (i) Increasing cache
efficiency of the fully explicit finite element procedure by
decomposition of the global data to smaller sub-blocks with
application of domain decomposition concept. (ii) Extension of the
introduced idea to parallel processing by application of secondary
inter-processor domain decomposition procedure. (iii) Introducing
easy to implement parallel explicit finite element procedure for
solution of related PDEs on the non-overlapped element-base
decomposed domain. (iv) Solution of some technical problems behind
suitable domain decomposition for practical casting simulation
(problems due to presence of decupled meshes, e.g., cast mesh is
completely decupled from mold mesh). (v) Application of high
performance computation for solution of a practical engineering
problem, i.e., casting solidification simulation.

The rest of this paper is organized as follows. In the next section,
the governing equations are presented. Section 3 contains the
numerical finite element approximation of the governing equations.
In Section 4 we present our cache efficient solver and details of
its implementation for serial and parallel execution. Section 5 is
devoted to experimental result  and section 6 summarizes the present
work.


\section{Mathematical modeling}

\noindent From a macroscopic point of view, if effect of melt flow
during solidification is neglected, solidification is governed by
the heat conduction equation. The appropriate mathematical
description of the heat conduction process in a stationary material
region, $\Omega$ is given by,
\begin{equation} \label{}
\rho c\ {\partial{T}\over {\partial t}}= \nabla\cdot\ (k \nabla T)
\end{equation}
where $T$ is the temperature field, $\rho$ is the density, $c$ is
the specific heat and $k$ is the thermal conductivity. The initial
condition is
\begin{equation} \label{}
T( t=0 ) = T_0
\end{equation}
where $T_0$ is pouring temperature of liquid metal. The boundary
condition used are given by:
\begin{equation} \label{}
T = f(x,y,z,t) \qquad \textrm{on} \quad \Gamma_{T}
\end{equation}
\begin{equation} \label{}
-k{\partial{T}\over {\partial n}}  = q + h ( T - T_{\infty} ) \qquad
\textrm{on} \quad \Gamma_{q}
\end{equation}
where $q$, $h$ and $T_{\infty}$ are specified boundary heat flux,
convective heat transfer coefficient and the ambient temperature
respectively and $n$ is the direction of the unit normal vector to
$\Gamma_{q}$. $\Gamma_{T}$ and $\Gamma_{q}$ are the portions of the
boundary $\Gamma$ of the computational domain $\Omega$.

The enthalpy method is one of the popular methods for incorporation
of phase change effect in the conductive heat transfer equation. in
this manner the equation (1) has the following form (Swaminathan and
Voller, 1992):
\begin{equation} \label{enthalpy}
{\partial{H}\over {\partial t}}= \nabla\cdot\ (k \nabla T)
\end{equation}
where $H$ is the enthalpy field. In general the enthalpy could be a
function of a number of variables, such as temperature,
concentration, cooling rate, etc. In many solidification model,
however, the enthalpy in the mushy region can be assumed to be a
function of temperature alone. There are several method for solution
of equation (\ref{enthalpy}) that, in the present study the apparent
heat capacity method (Comini et al., 1974) is used as follow:
\begin{equation} \label{cpa}
{\partial{H}\over {\partial T}}{\partial{T}\over {\partial t}}=
\nabla\cdot\ (k \nabla T)
\end{equation}
\begin{equation} \label{}
C_{P}^{app}={\partial{H}\over {\partial T}}= \bigg(
      {{\left(\partial{H}/{\partial x}\right)^2 +
       \left(\partial{H}/{\partial y}\right)^2 +
       \left(\partial{H}/{\partial z}\right)^2  } \over
      {\left(\partial{T}/{\partial x}\right)^2 +
       \left(\partial{T}/{\partial y}\right)^2 +
       \left(\partial{T}/{\partial z}\right)^2  }}
\bigg)^{1/2}
\end{equation}
\newline
$C_{P}^{app}$ in equation (\ref{cpa}) called apparent heat capacity,
that is evaluated by Lemmon (Lemmon, 1981) method in this study.
This method shows good robustness and accuracy (Thomas et al.,
1984).

An essential step in the accurate simulation of the cooling of a
cast part is to model the effect of air gap formation between the
mould and casting and other contact resistance between two
components of casting system. The problem of contact resistance
between two stationary materials may be represented by either of two
methods. One approach to modeling this effect assumes that the
contact region is composed of a fictitious material, of small
thickness, whose properties produce the appropriate resistance to
heat flow across the interface. Typically, this gap material will
have a negligible heat capacity and a nonlinear conductivity. A more
convenient model can be introduced if the interface contact is seen
as a convective thermal resistance between two surfaces. In the
other words, one can easily replace the concept of contact
conductance with an equivalent convection heat transfer coefficient.
The model can also accommodate radiation heat transfer between
surfaces of the mould and casting. In this model the geometry,
including the casting and mould, is separated by an interface
boundary line in the two-dimensional and a surface in the
three-dimensional case. The interface boundary is subdivided into a
number of segments that represent different interface conditions.
Each boundary segment is identified by two interface segments; a
segment which belongs to the casting region and a corresponding
segment associated with the mould region. These two segments are
actually lying on each other and represent the same boundary. For
each boundary segment in the casting, the corresponding boundary
segment in the mould acts as an ambient condition and vice versa.
This means that the heat transfer between these two surfaces is
characterized by their temperatures and the local heat transfer
coefficient specified for that segment of the interface:
\begin{equation} \label{}
-k{\partial{T^{cast}}\over {\partial n}}\vert_{interface} =
 k{\partial{T^{mold}}\over {\partial n}}\vert_{interface} =
 h_i ( T^{cast} - T^{mold} )\vert_{interface}
\end{equation}
where $h_i$ is local heat transfer coefficient of interface. the
value of $h_i$ must be extracted from material data base. It can be
fixed or be function of time and temperature. The concept of
interfacial convective heat transfer can easily be applied to any
kind of contact surface in casting configuration. This might be a
surface between a chill and casting or mould and a surface
separating two parts of a mould as shown in Figure 1. For instance,
$A_{cast}$ and $A_{cope}$ represent interfacial boundaries of the
cast and cope for segment $A$. In this way, virtually all contact
surfaces in the casting system can be conveniently modeled (Manzari
et al., 2000). The other benefits of this approach are:
\begin{enumerate}
\item[(i)]
It is possible to generate mesh separately in cast and mold and
don't need matching element in the interface (see Figure 2). This
gives more flexibility to mesh generator and lead to better mesh
quality and also increase mesh generation efficiency.
\item[(ii)]
The element sizes for two sides of the interface can be
significantly different. This is often useful as a relatively dense
mesh is normally employed for the casting region in comparison with
the mould region.
\item[(iii)]
There is no need to define the thermal conductivity and air gap
thickness as they are uniquely replaced by the interfacial heat
transfer coefficient.
\item[(iv)]
Any standard finite element program can be used for this purpose and
no modification is required.
\item[(v)]
This model will help to assure a smoother temperature distribution
when starting with a large temperature difference between two sides
of the interface.
\item[(vi)]
The ability to accommodate both spatial and temporal variations in
the interfacial heat transfer coefficients.
\end{enumerate}

\noindent For more detail about the solidification modeling see
Refs. Huy and Argyropoulos (1996) and Lewis and Ravindran (2000).

\begin{figure}[ht]
\begin{center}
\includegraphics[width=15.cm,height=6.cm]{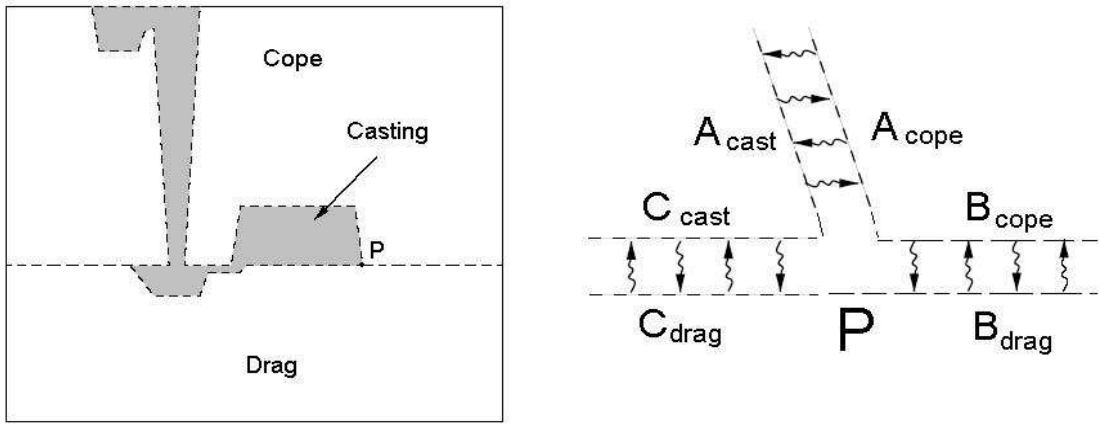}
\caption{Separation of regions at interfaces and assignment of the
associated boundary conditions (this figure is adapted from a
presentation by Prof. Mehrdad T. Manzari).}
\end{center}
\end{figure}

\begin{figure}[h]
\begin{center}
\includegraphics[width=8.cm,height=5.cm]{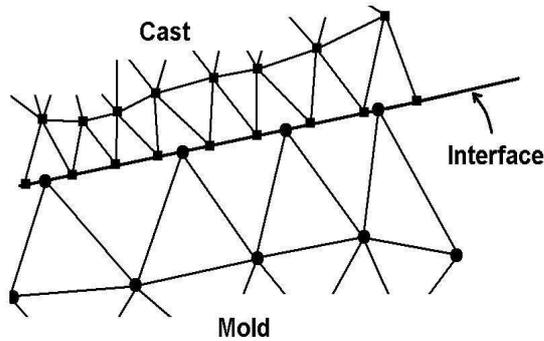}
\caption{Schematic of non-matching elements coordination for a
typical cast-mold interface.}
\end{center}
\end{figure}


\section{Numerical modeling}

\noindent For solution of the governing equations we subdivide
spatial physical domain with unstructured four node tetrahedral
elements. The standard Galerkin form of the weighted residual method
is used to discretize equation (\ref{cpa}) in space by the FEM,
which yields the following non-linear system first order
differential equations a system of first-order ordinary differential
equations:
\begin{equation} \label{nheat}
C(T){\partial{T}\over {\partial t}}+ K(T)T = F(T)
\end{equation}
where $C$ and $K$ are respectively capacitance conductance matrices,
vector T is the vector of nodal unknowns and the right–hand side
vector F contains boundary fluxes. The mass lumping method
(diagonalizing capacitance matrix with 'row-sum' technique) is used
for calculation of capacitance matrix. It has better stability and
computational efficiency and also in our parallel implementation
needs lower communications cost. The traditional finite difference
method is generally used to obtain a fully discretised algebraic
system of equations for the time domain as follows (Lewis et al.,
1996):
\begin{equation} \label{teta}
\big( \ {C \over \delta t} - \theta K \ \big)\ T^{n+1} = \big( \ {C
\over \delta t} - ( 1 - \theta )\ K \ \big)\ T^n + F^n
\end{equation}
where $\delta t$ is time step and $\theta$ is blending factor for
switching between explicit and implicit method. First order fully
explicit and implicit method is achieved with $\theta$ equal to 0
and 1 respectively and $\theta=0.5$ give implicit second order
accurate Crank-Nicholson method. The nodal values of enthalpy are
related to the temperature by:
\begin{equation} \label{}
H = \int_{e} C_{P}^{app}\mathrm{d} T
\end{equation}
where $H$ is the enthalpy per unit mass (variation of density is
neglected in this study). The temperature field is then obtained by
solving such a set of equations. Because of the apparent heat
capacity and thermal conductivity are function of temperature field,
equation (\ref{nheat}) is non-linear. In this study, equation
(\ref{nheat}) is linearized by using known values of apparent heat
capacity and thermal conductivity (from previous step) and done
iterative updating procedure until achieving convergence as follows:
\begin{equation} \label{}
\big( \ {C(T^m) \over \delta t} - \theta K(T^m) \ \big)\ T^{m+1} =
\big( \ {C(T^m) \over \delta t} - ( 1 - \theta )\ K(T^m) \ \big)\
T^n + F^n
\end{equation}
where superscript $m$ is related to non-linear iteration. After
convergence of non-linear iteration, last values of temperature
field is used as $T^{n+1}$. In the explicit time integration method,
by using enough small time step which, is usually stability bound of
explicit method, non-linear loop is not essential (for more detail
see Lewis et al. (1996) and Reddy and Gartling (2000)). The
stability bound of the explicit time integration method is
approximately given by,
\begin{equation} \label{}
 \delta t  < \frac{\rho c l^2}{k}
\end{equation}
where $\rho$ is density, $c$ is the nodal heat capacity, $k$ is the
nodal heat conductivity and $l$ is the minimum mesh distance.
Although the implicit time integration method is unconditionally
stable but increasing time step size leads to decreasing overall
time accuracy and slow convergence rate (or divergence) of
non-linear iterations.

As mentioned in previous section the heat transfer between cast and
mold is replaced by equivalent convective heat flux. The ambient
temperature is found for each interface element by using the average
temperature of the its closest element in the other region. Therfore
at start stage of simulation closest element of each interface
element (in the other region) is found and is labeled as sister of
it for interpolation of ambient temperature during simulation.


\section {Implementation of cache efficient solver}

Since the parallel implementation of solver is a part of this study
and it has complete analogy with implementation of cache efficient
algorithm we present method of parallel implementation first and
then describe our cache efficient solver.

\subsection {Parallel implementation\\}
\subsubsection {Domain decomposition}
\noindent The first aspect of parallelizing a finite element method
by domain decomposition is to divide the grid into parts. This is
how the full computational task is divided among the various
processors of the parallel machine. Each processor works only on its
specific portion of the grid and anytime a processor needs
information from another processor a message is passed.

For the best parallel performance, one would like to have optimal
load balancing and as little communication between processors as
possible. Consider load balancing first. Ideally, one would like
each processor to do exactly the same amount of work every
iteration. That way, each processor is always working and not
sitting idle. For a finite element code, the basic computational
unit usually is the element. It makes sense to partition the grid
such that each processor gets an equal (or nearly equal) number of
elements to work on. The second criterion is that the amount of
communication between processors be made as small as possible. This
needs minimizing number of edge cuts and the degree of adjacency
(number of neighbors) for each processor. For partitioning procedure
we use the METIS library (Karypis and Kumar, 1998; Karypis and
Kumar, 2005)  that done first criteria excellently and second
criteria in the acceptable manner.

As mentioned in previous section, mesh generation of each component
of casting configuration is done separately and in contact surfaces
there aren't essentially coincidence of grid points. Therefore we
have two (or more) separated grids. This discontinuity in global
grid, decrease quality of mesh partitioning, i.e., increase
probability of formation of low quality and separated sub-domains.
It is natural result when the two or more meshes without any
physical connection are undergone domain decomposition
simultaneously.

This problem is solved in this study by adding some virtual elements
between each two contact surfaces, i.e., for each cast surface
element the corresponding sister element is indicated and by its
three nodes and one nodes of its sister element, a virtual
tetrahedron is formed and added to end of element connectivity list
(see Figure 3). These virtual elements are only used for domain
decomposition purpose.  For improvement of this algorithm it is
better to add such virtual elements only in one side of two regions
for decreasing the number of sub-domains that are shared in two
regions (cast and mold). Figure 4 shows the simple casting system
that has two separated mesh (cast and mold) which is divided to six
sub-domains by METIS library. The left section of Figure 4 shows
result of traditional partitioning and right section, shows result
of presented algorithm (adding virtual element to left side of
cast-mold interface). The number of edge cuts for this sample are:
6539 and 1625, for traditional partitioning and presented algorithm
respectively. Also the load balancing of the presented domain
decomposition method is about $10 \%$ better than the primary
method. It can be seen that by using presented algorithm, the
quality of partitioning procedure is increased considerably.

The non-overlapped element based domain decomposition method is used
for parallelizing finite element procedure in the present study. In
this manner the elements are assigned uniquely to partitions.
Therefore after domain decomposition each processor has own private
nodes and shared nodes with other processors in its boundaries with
other sub-domains. If boundary between two neighbor sub-domains is
cast-mold interface (or in general case interface of two casting
components), we have additional kind of nodes that we called them as
external nodes (nodes related to sister elements) as shown in Figure
5.

\begin{figure}
\begin{center}
\includegraphics[width=8.cm,height=6.cm]{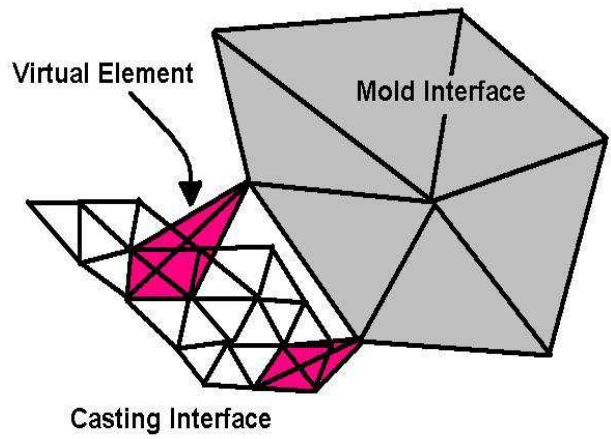}
\caption{Schematic of addition of virtual element to cast-mold
interface.}
\end{center}
\end{figure}

\begin{figure}
\begin{center}
\includegraphics[width=16.cm,height=8.cm]{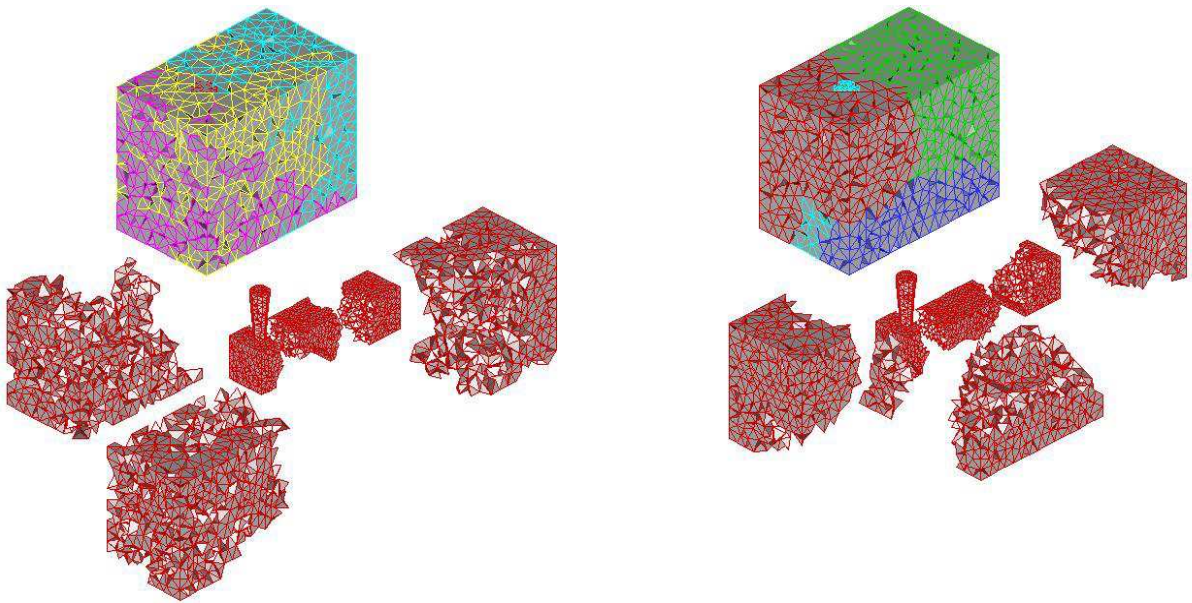}
\caption{Comparison between result of traditional domain
decomposition (left) and adding virtual contact element (right).}
\end{center}
\end{figure}

\subsubsection {Parallel implementation of explicit solver}

\noindent When $\theta=0$ in equation (\ref{teta}), the pure
explicit algorithm is preserved as follows:
\begin{equation} \label{update}
T^{n+1} = T^{n} + {\delta t}\ {\big(C^n\big)}^{-1} \  \big(F^n - K^n
T^n  \big)
\end{equation}
For calculation of new temperature field, the global capacitance and
conductance matrix must be calculated. Because of we use mass
lumping method matrix $C$ is diagonal, therefore updating procedure
of each node contains one matrix-vector product and two vector
summation and one vector scaling.

In a distributed environment, calculation of $C^n$ and expression $\
F^n - K^n T^n$ requires communication between processors. The
capacitance matrix, $C$ (is diagonal), the conductivity matrix, $K$,
and boundary flux vector, $F$ can be written as
\begin{equation}\label{}
K \ = \sum_{e=1}^{N} \ {\hat{K}}^e
\end{equation}
\begin{equation}\label{}
C \ = \sum_{e=1}^{N} \ {\hat{C}}^e
\end{equation}
\begin{equation}\label{}
F \ = \sum_{e=1}^{N} \ {\hat{F}}^e
\end{equation}
where ${\hat{C}}^e$, ${\hat{K}}^e$ and ${\hat{F}}^e$ are the
contribution from the element $e$ and $N$ is the total number of
elements. ${\hat{K}}^e$ and ${\hat{F}}^e$ represent the local
element contribution ($K^e$ and $F^e$ respectively) expanded to
system size, i.e., the entries of ${\hat{K}}^e$ are mapped into
corresponding global row and column locations with others entries in
${\hat{K}}^e$ equal to zero. Then  equation (\ref{update}) can be
written as
\begin{eqnarray} \label{update2}
\lefteqn{} \nonumber\\ \lefteqn{T^{n+1} = T^{n} + {\delta t}\
{\big(C^n\big)}^{-1} \  \big(F^n - K^n T^n  \big)} \nonumber\\ & &
{} {} \nonumber\\ & & {} {\ \ = T^{n} + {\delta t}\ {\Bigg(
\sum_{e=1}^{N} \ {\hat{C}}^e \Bigg)}^{-1} \  \ \Bigg[ \Bigg(
\sum_{e=1}^{N} \ {\hat{F}}^e \Bigg) - \Bigg( \sum_{e=1}^{N} \
{\hat{K}}^e \Bigg)\ T \Bigg]}
\end{eqnarray}
${\hat{K}}^e$ is a very sparse matrix, but only the dense local
element contribution $K^e$ needs to be stored. $K^e$ and $F^e$ can
be calculated concurrently and independently for elements
$e=1,2,…,N$. We define the local capacitance matrix ( $  C^{l}_i  $
), the local conductivity matrix ( $  K^{l}_i  $ )  and the local
boundary flux vector ( $ F^{l}_i $ ) of each processor as follows
\begin{equation}\label{}
C^{l}_i \ = \sum_{e=1}^{N_l} \ {\hat{C}}^e
\end{equation}
\begin{equation}\label{}
K^{l}_i \ = \sum_{e=1}^{N_l} \ {\hat{K}}^e
\end{equation}
\begin{equation}\label{}
F^{l}_i \ = \sum_{e=1}^{N_l} \ {\hat{F}}^e
\end{equation}
where $N_l$ is the number of element in each partition (related to
each processor). So we have
\begin{eqnarray} \label{update3}
\lefteqn{T^{n+1} = T^{n} + {\delta t}\ {\Bigg( \sum_{i=1}^{N_p} \
{C}^l_i \Bigg)}^{-1} \  \ \Bigg[ \Bigg( \sum_{i=1}^{N_p} \ {F}^l_i
\Bigg) - \Bigg( \sum_{i=1}^{N_p} \ {K}^l_i \Bigg)\ T \Bigg]}
\nonumber\\ \lefteqn{} \nonumber\\ & & {} {} \nonumber\\ & & {} {\ \
= T^{n} + {\delta t}\ {\Bigg( \sum_{i=1}^{N_p} \ {C}^l_i
\Bigg)}^{-1} \  \ \Bigg[
 \sum_{i=1}^{N_p} \ \bigg({F}^l_i  - {K}^l_i \ {T}^l_i \bigg)\
\Bigg]}
\end{eqnarray}
where $N_p$ is number of partitions and ${T}^l_i$ is local
temperature vector of ith-partition. Note that each partition must
have the global values of vector $T^n$ in its shared nodes for
performing above procedure.

Updating procedure in this study is as follows: Using traditional
element-by-element finite element procedure and calculation of
${C}^l_i$ and expression $\sum_{i=1}^{N_b} \ \big({F}^l_i  - {K}^l_i
\ {T}^l_i \big)$, locally within each partition. For internal nodes
the above procedure is complete and equation (\ref{update3}) can be
used for updating temperature field.  But any shared node, needs
effects of elements that constructed from it, but posed in the
neighbor partitions. These effects are equal to calculated values of
${C}^l_i$ and expression $\sum_{i=1}^{N_b} \ \big({F}^l_i  - {K}^l_i
\ {T}^l_i \big)$ by neighbor partitions. Therefore neighbor
partitions must exchange these calculated values related to their
shared nodes and perform summation on exchanged data. Note that for
imposition of boundary condition each block needs to temperature
values of its external nodes (see Figure 5) that achieves them via
communication. After this step, new temperature field is calculated
with equation (\ref{update3}).

The size of data that must be communicated between two neighbor
processors is equal to, two times of number of their shared nodes.
The partitions that have mold-cast interface as inter partition
boundary, need additional communication for obtaining temperature of
theire external nodes.

Note that main difference of our parallel solver with related works
are non-overlapped domain decomposition and communication of
calculated values vs. overlapped domain decomposition and
communication of temperature values of overlapped region.

\begin{figure}
\begin{center}
\includegraphics[width=16.cm,height=7.cm]{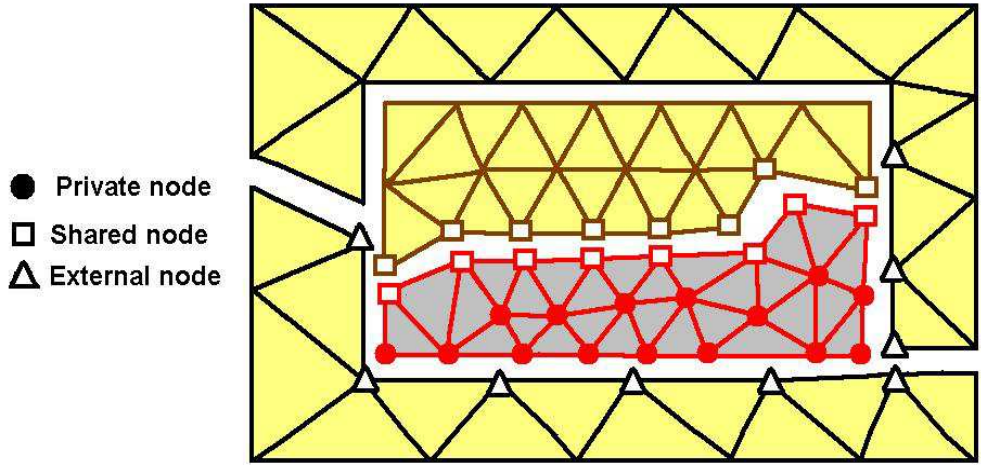}
\caption{Element based domain decomposition and definition of
various kind of nodes that are used in this study.}
\end{center}
\end{figure}

\subsubsection {Communication scheduling}

\noindent We need for a communication schedule to coordinate the
exchange of information between processors. A communication schedule
simply tells each processor when and with whom it can exchange its
message. This is necessary because a processor can only communicate
with one other processor at a time.

Consider the communication graph in Figure. 6. Each node represent
one processor and each edge shows needful communication. In this
case there are 4 separate communication stages, and in all but the
third, two separate pairs of processors swap information at the same
time. For example, in stage 2, processors 1 and 3 swap as do
processors 2 and 0. Processor 4 is idle. So you can see that it
isn't necessary to have a separate communication stage for each pair
of processors that need to communicate. It is only necessary to make
sure that pairs of processors swap without conflict.

This communication schedule is constructed by coloring the edges of
the graph so that each edge coming from a vertex has a different
color. An attempt is made to minimize the number of colors used
which equals the number of communication stages that must be
performed. Here is an outline of the algorithm:
\begin{enumerate}
\item[(i)]
Determine the vertex of highest degree. Call it the root
\item[(ii)]
Starting at the root look at all the edges emanating from the vertex
and determine which ones have not yet been colored.
\item[(iii)]
Mark all these uncolored edges with a different color starting with
the lowest available color and working up.
\item[(iv)]
Do this same procedure for all the rest of the vertices in the graph
until all the edges are colored.
\end{enumerate}

There is a theorem in graph theory called Vizings theorem which
states that the most colors one would ever have to use for this is
equal to the degree of the root plus one. The algorithm presented
above has never produced more colors than dictated by Vizings
theorem.

\begin{figure}
\begin{center}
\includegraphics[width=5.cm,height=5.cm]{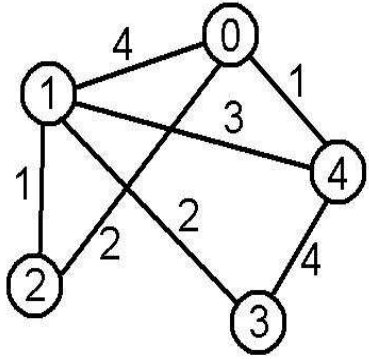}
\caption{A sample of communication graph.}
\end{center}
\end{figure}

\subsubsection{Message passing interface}

\noindent We use MPI library for the message passing implementation
of the exchange algorithm. In this section we describe simple
blocking communication method that is used in this study.

Using blocking massage passing needs care for prevention of decaying
efficiency and system dead-lock. Scheduling communication table
(that is described in previous section) is essential for efficient
implementation of blocking communication (although with non-blocking
communication, scheduling is not essential, but fore controlling
network traffic and consequently achieving better performance
scheduling is recommended). For this purpose we perform
communication based on table of communication with function
MPI\_Sendrecv. This call performs a data swap, by combining the send
and receive into a single call. The communication with this routine
is faster than doing a separate synchronous send and synchronous
receive. The other benefit is prevention of system dead lock that
may be happen during blocking communication.

\subsection{Improvement cache efficiency}

\noindent As stated in the introduction section, the key idea behind
optimization of cache behavior is improvement data locality with
decomposition of computation into smaller parts (cache blocks) and
doing all of the computing in each cache block one at a time. For
this purpose we use from domain decomposition concept. So we
decompose the global mesh to appropriate number of cache blocks and
treat each block independently from others and anytime a block needs
information from another blocks communicate (virtually) with them.
This algorithm is similar with our parallel implementation. The main
difference is that in the parallel solver the communication between
neighbor partitions is performed via network (intra-processor
communication), while in the cache efficient algorithm,
communication between neighbor blocks is performed directly
(inter-processor virtual communication). Therefore the domain
decomposition, solver implementation and essential data structure
are same as parallel solver.

This method can be easily extended to parallel processing. For this
purpose, after primary domain decomposition based on load balancing
criteria and division of global mesh between processors, each
processor performs the secondary domain decomposition for
construction of its cache blocks. Figure 7 shows schematically this
procedure.

Optimal number of cache block is severely function of cache size,
specially L2-cache. In the present study this parameter is indicated
with simple numerical experiment. For this purpose each processor
changes number of its blocks and solve problem locally for few
number of time steps (without communication with other processors).
Then the optimal number of blocks for each processor is indicated
based on the measured elapsed CPU time. As we deal with transient
problem the computational overhead of this stage is negligible
(below 1 \% of total time for practical application in our
experiment).

The other advantage of this algorithm (which is not in the scope of
present study) is providing possibility of multi-thread execution on
each processor (executing computation of each block or each some
blocks as one thread). This is very important for high performance
computation when we deal with hyper-threaded technology CPUs or
dual-processor CPUs. For such systems single thread execution can
take only 50 \% of CPU's power in the extreme case. But with
multi-threading it is possible to take nearly 100 \% of CPU usage.

\begin{figure}[th]
\begin{center}
\includegraphics[width=15.cm,height=8.cm]{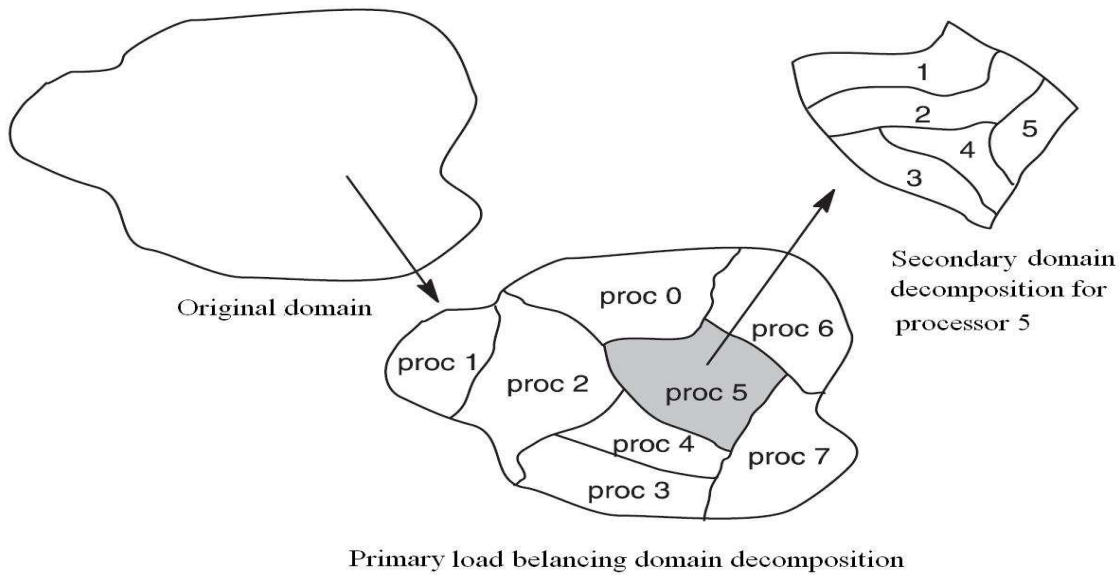}
\caption{Two level domain decomposition method: original domain is
first partitioned based on load balancing criteria and divided
between processors, and then each processor divides its local mesh
to appropriate number of block for improving cache efficiency, this
figure is adapted from (Gullerud and Dodds).}
\end{center}
\end{figure}


\section {Result and disscution}

\noindent In this section we investigate performance of presented
algorithm by solution of the problem for six test cases with various
mesh resolution and configuration. In these cases grids are
generated individually over casting and mold part without attention
to matching grids at cast-mold interfaces. Figure 8  shows the
surface mesh (cases 1-4) and Table 2 shows the mesh properties of
these cases.

\begin{figure}
\begin{center}
\includegraphics[width=8.cm,height=20.cm]{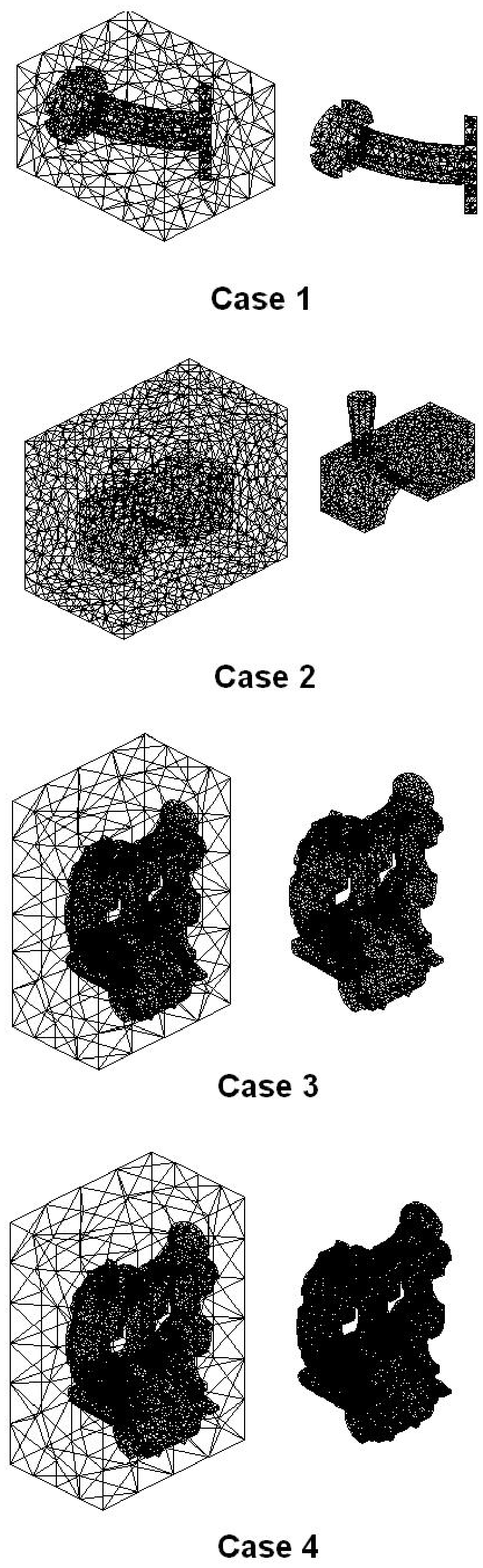}
\caption{Surface mesh configuration of cases 1-4 in this study.}
\end{center}
\end{figure}

\subsection{Computational resource}

\noindent As the computing platform for parallel execution, we have
used a 14 nodes Beowulf cluster based on 900 MHz Intel Pentium III
processors and switched Ethernet networks (100Mbps). Table 1 shows
specification of various machines which were used for serial
execution in the present study.

\subsection {Performance of mesh reordering}

\noindent In this experiment we investigate effect of mesh
reordering with reverse Cuthill-McKee method (Alan and Liu, 1981) on
performance of serial code. Figure 9 shows the sparsity pattern of
global coefficient matrix befor and after mesh reordering and Table
3 shows the effect of reordering on the bandwidth of global
coefficient matrix. It is clear that the bandwidth of global
coefficient matrix is reduced considerably with reordering
procedure. Table 4 shows the effect of reordering on performance of
sequential explicit finite element procedure for cases 1-6 on
various machines. It can be seen that the performance of computation
is increased with this procedure and this effect is increased with
decreasing size of the global data.

\subsection {Performance of the serial cache efficient algorithm }

\noindent In this section we present results of application of the
presented method on the efficiency of sequential explicit finite
element procedure. We define the performance improvement factor as
relation of CPU time of original explicit finite element procedure
to CPU time of presented cache efficient finite element procedure.

Figures 10 - 15 show the variation of improvement factor with
variation of number of blocks for cases 1-6 for non-reordered (a)
and reverse Cuthill-McKee reordered (b) mesh. Plots show that with
increasing number of blocks the performance is increased and a
performance peak is achieved in the appropriate number of blocks.
The performance improvement is generally better for the large
meshes. This is natural result because with increasing size of data
the probability of cache miss is increased due to increasing data
sparsity. Also it seems that for relatively small data (cases 1-3)
the performance peak of reordered mesh is lower than non-reordered
mesh. But in spite of this smaller performance peak in the all of
our experiment the CPU time related to reordered mesh is generally
better than non-reordered mesh. So combination of reordering and
cache blocking leads to better performance. Note that with
increasing number of blocks the total number of nodes of global data
is increased considerably  (due to presence of shared nodes) as
illustrated in Figures 10 (c) - 15 (c). It is equivalent to
increasing size of the global data. But in spite of this the overall
effect leads to better performance (total number of elements is
fixed). The average size of blocks are also plotted in 11 (d) - 16
(d).

Figure 16 shows the average block size of cases 1-6 at the
performance peak for each machine. Plot shows that the optimal block
size is approximately equal to size of L2-cache (except case 1 that
its data is small and increasing number of blocks severely increase
the size of global data due to presence of shared nodes).

\subsection {Performance of the parallel cache efficient algorithm }

\noindent In this section we present result of parallel version of
the presented method. Figure 17 shows the speedup curves (for cases
1-6) related to parallel execution of the presented method (original
and cache efficient). Note that the CPU time of serial execution of
the traditional method was used as reference CPU time for
calculation of speedup. Plot shows that for original method good
parallel efficiency and scalability is resulted and in spite of
application of the Ethernet network, nearly linear speedup is
achieved for relatively heavy meshes (cases 3-6). Also the presented
cache efficient algorithm preserves its efficiency in the parallel
execution and the efficiency is improved with factor 1.2-2.2 (except
case 1).


\section {Conclusions}

\noindent In this paper, a simple method is presented for
improvement cache efficiency of the explicit finite element
procedure with application to casting solidification simulation
(non-liner heat transfer problem with phase change) over
three-dimensional complex geometries. In this method the global data
is divided to smaller blocks with traditional domain decomposition
method and calculation related to each block is performed
independently from others. A novel algorithm is presented for
solution of the governing equations on the non-overlapped
element-base decomposed domain. Then the presented method is
extended to parallel processing without considerable additional
effort. The other covered features are: presentation of a simple
algorithm for improvement quality of the decomposed domain when the
original global mesh contains two or more decoupled meshes;
investigation effect of mesh reordering on the performance of the
explicit finite element procedure; proposing efficient
inter-processor data communication based on communication scheduling
and blocking-mode massage passing.

Our results show that: 10-20 \% performance improvement is achieved
with mesh reordering; achieving nearly linear speedup with combining
of the presented parallel finite element procedure and data
communication method on a Ethernet networked distributed memory
parallel machine; application of the presented method leads to
increasing computational efficiency of serial and parallel execution
with factor 1.2-2.2; the optimal block size is approximately equal
to size of the L2-cache.


\vspace{10mm} {\Large\bf\noindent Acknowledgements\\} \noindent This
article was the term project (number 2) of my parallel processing
course due to Prof. Mehrdad T. Manzari. I would like to thanks Dr.
M.T. Manzari for helpful discussion and sharing his sequential FEM
code with me to parallelize.




\begin{table}
\caption{Specification of various machines, which were used for
serial execution in the present study.} \footnotesize\rm
\begin{tabular*}{\textwidth}{c c c c c c c c c}
\\\hline\\
Machine&&CPU (MHz)&&L1-cache (Kb)&&L2-cache (Kb)&&RAM (Mb)\\
\\\hline\\
Machine 1&&AMD 2000&&128&&512&&512\\\\
Machine 2&&AMD 1000&&128&&64&&256\\\\
Machine 3&&Intel 2400&&8&&512&&512\\\\
Machine 4&&Intel 3200&&15&&1024&&512\\\\
Machine 5&&Intel 3000&&8&&512&&512\\\\
\hline\\
\\\\
\end{tabular*}
\end{table}

\begin{table}
\caption{Mesh properties for cases 1-6 in the present study.}
\footnotesize\rm
\begin{tabular*}{\textwidth}{c c l c l}
\\\hline\\Case&&Number of Nodes&&Number of Tetraheders \\
\\\hline\\
\ 1&& 348&&7638\\\\
\ 2&&4065&&18388\\\\
\ 3&&14800&&64989\\\\
\ 4&&29205&&131582\\\\
\ 5&&109119&&540280\\\\
\ 6&&149657&&741690\\\\
\hline\\
\end{tabular*}
\end{table}

\begin{table}
\caption{Bandwidth of global coefficient matrix for cases 1-6 in the
present study, before and after mesh reordering with reverse
Cuthill-McKee method.} \footnotesize\rm
\begin{tabular*}{\textwidth}{c c l c l}
\\\hline\\
Case&&Initial Bandwidth&&Bandwidth after reordering\\
\\\hline\\
\ 1&&412&&266\\\\
\ 2&&2135&&254\\\\
\ 3&&14694&&2285\\\\
\ 4&&27631&&4264\\\\
\ 5&&108610&&1791\\\\
\ 6&&148658&&2358\\\\
\hline\\
\\\\
\end{tabular*}
\end{table}

\begin{table}
\caption{Effect of reverse Cuthill-McKee mesh reordering on
performance of sequential finite element procedure (performance
improvement in percent).} \footnotesize\rm
\begin{tabular*}{\textwidth}{c c c c c c c c c c c}
\\\hline\\
Case&&Machine 1&&Machine 2&&Machine 3&&Machine 4&&Machine 5\\
\\\hline\\
1&&7.6 \%&& 45.8\%&&5.3\%&&60.5\%&&37.1\%\\\\
2&&1.2\%&&42.2\%&&3.7\%&&48.9 \%&&36.1\%\\\\
3&&3.5\%&&33.5\%&&1.3\%&&6.9\%&&1.5\%\\\\
4&&1.\%&&30.3\%&&1.8\%&&18.2\%&&1.\%\\\\
5&&2.3\%&&21.7\%&&1.9\%&&10.0\%&&4.2\%\\\\
6&&3.0\%&&17.4\%&&5.1\%&&11.5\%&&7.9\%\\\\
\hline\\
\\\\
\end{tabular*}
\end{table}



\begin{figure}
\begin{center}
\includegraphics[width=7.cm,height=20.cm]{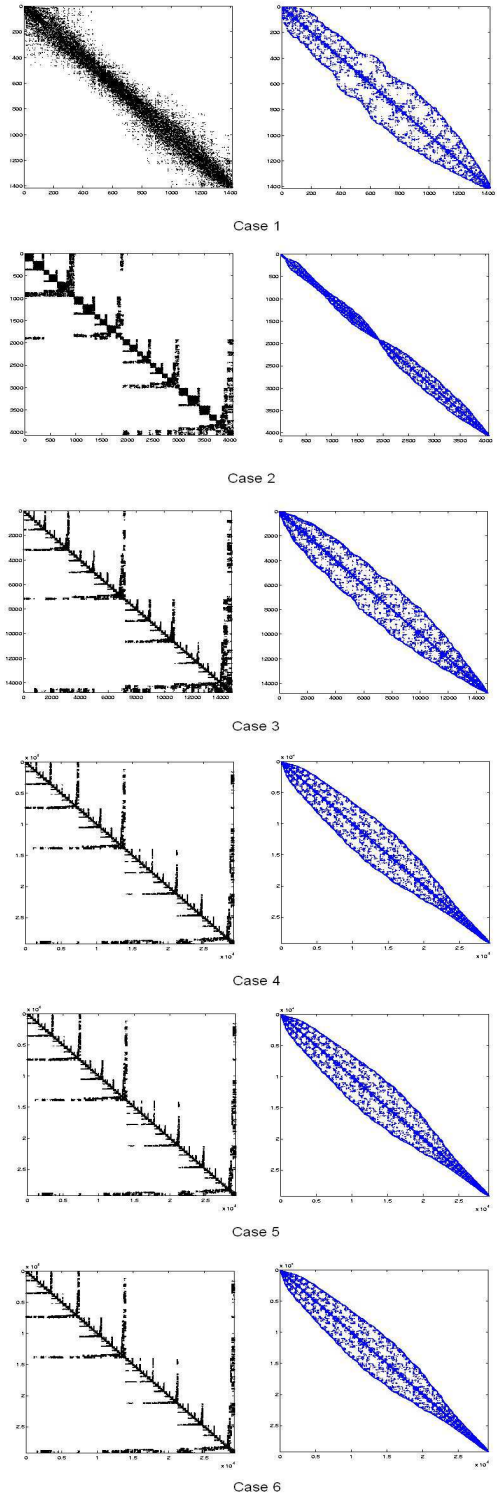}
\caption{sparsity pattern of global coefficient matrix for cases 1-6
in the present study, before (left) and after reordering with
reverse Cuthill-McKee reordering method (right).} \label{fig}
\end{center}
\end{figure}

\begin{figure}
\begin{center}
\includegraphics[width=15.cm,height=11.cm]{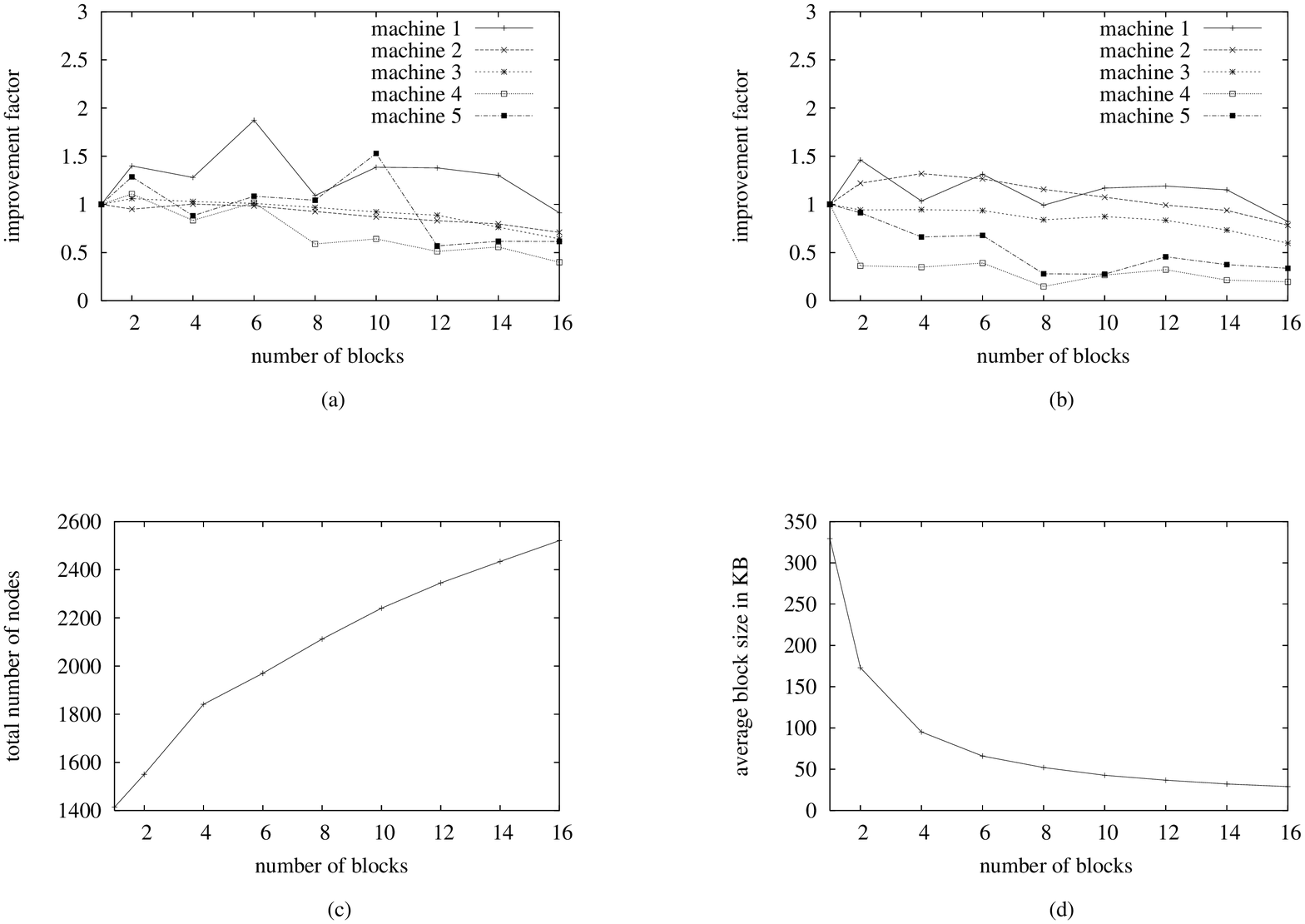}
\caption{Effect of variation of number of blocks on (a) performance
improvement factor of original mesh, (b) performance improvement
factor of reordered mesh, (c) total number of nodes and (d) averaged
block size for case 1.} \label{fig}
\end{center}
\end{figure}

\begin{figure}
\begin{center}
\includegraphics[width=15.cm,height=11.cm]{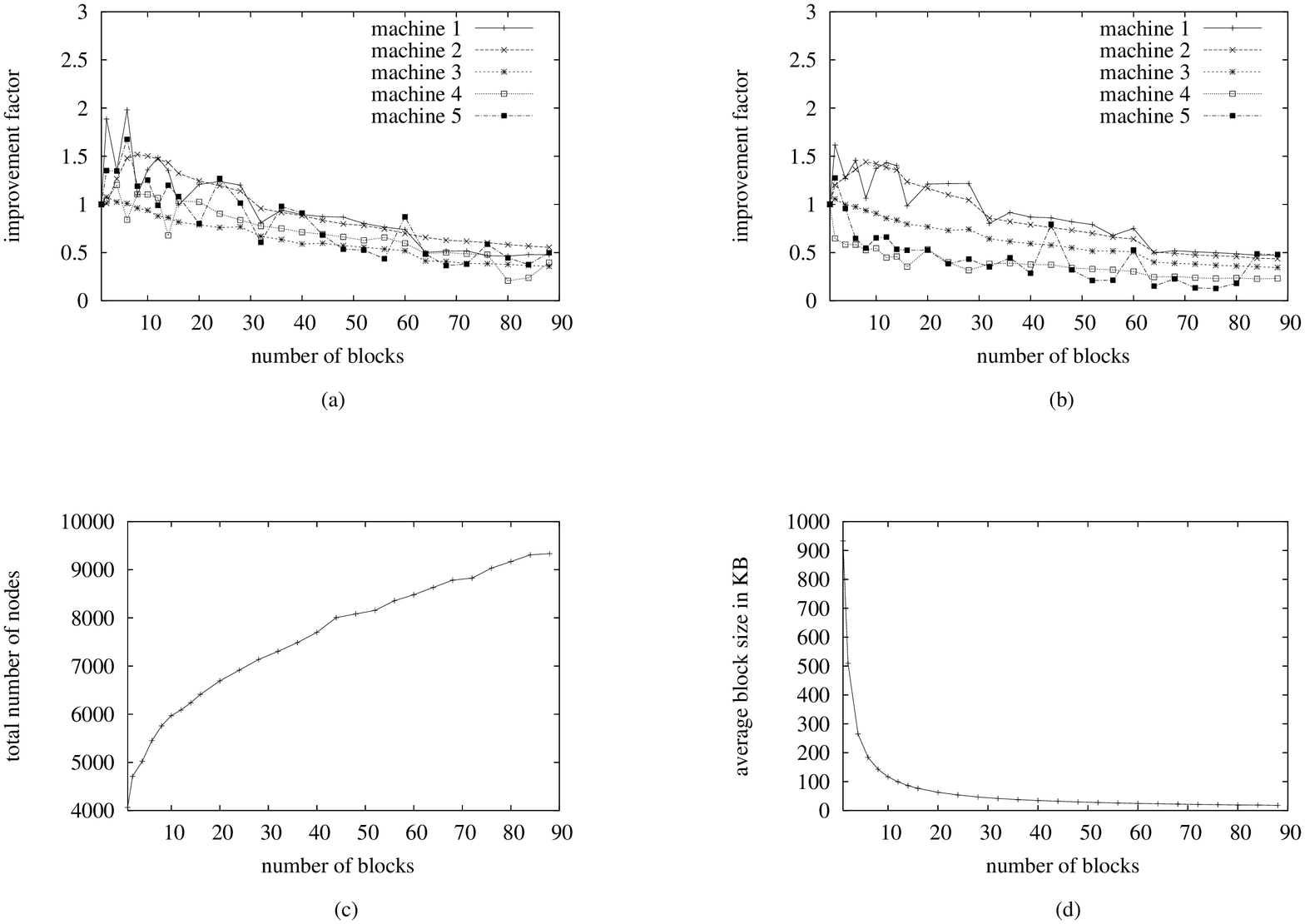}
\caption{Effect of variation of number of blocks on (a) performance
improvement factor of original mesh, (b) performance improvement
factor of reordered mesh, (c) total number of nodes and (d) averaged
block size for case 2.} \label{fig}
\end{center}
\end{figure}
\begin{center} \newpage \end{center}

\begin{figure}
\begin{center}
\includegraphics[width=15.cm,height=11.cm]{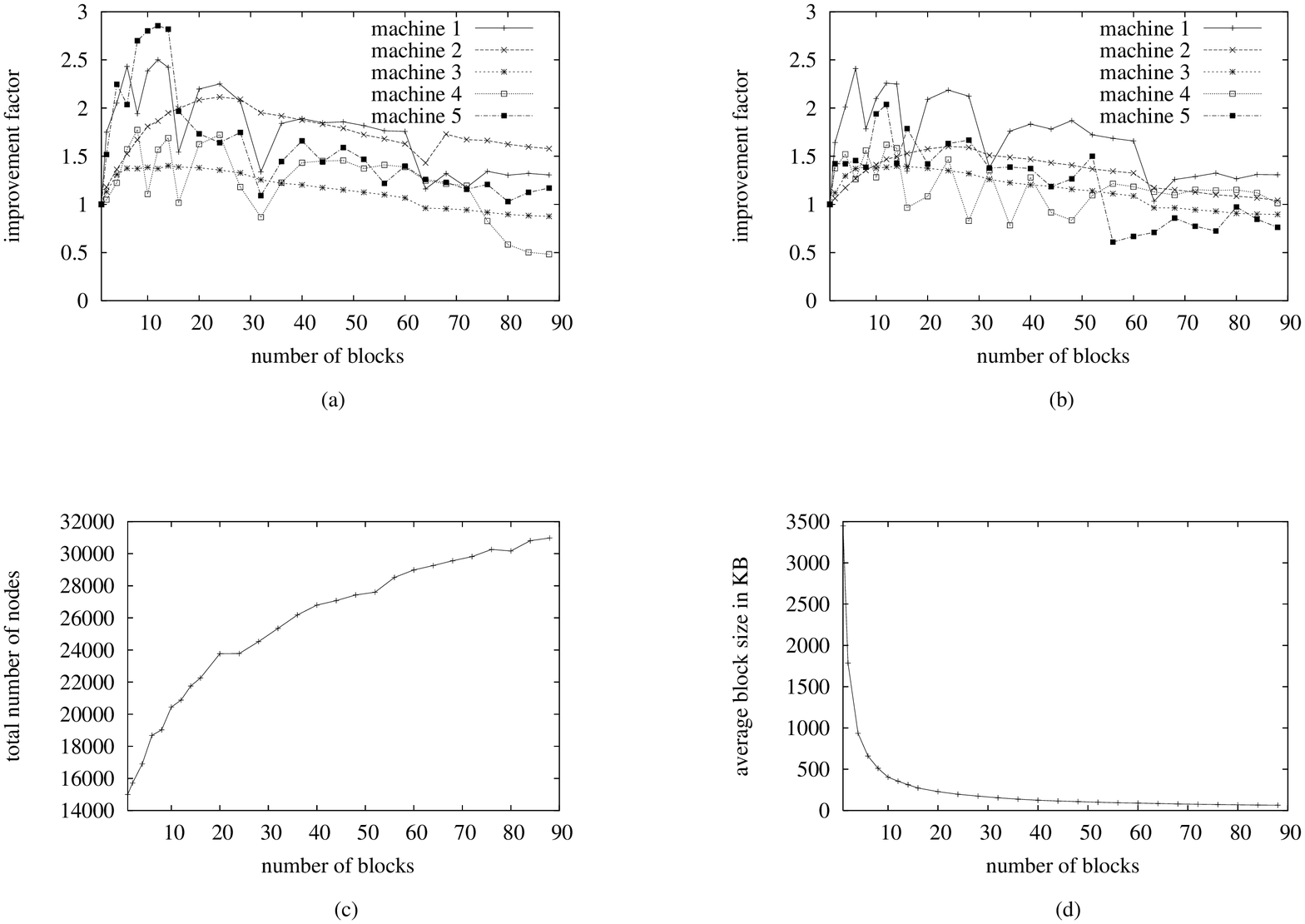}
\caption{Effect of variation of number of blocks on (a) performance
improvement factor of original mesh, (b) performance improvement
factor of reordered mesh, (c) total number of nodes and (d) averaged
block size for case 3.} \label{fig}
\end{center}
\end{figure}

\begin{figure}
\begin{center}
\includegraphics[width=15.cm,height=11.cm]{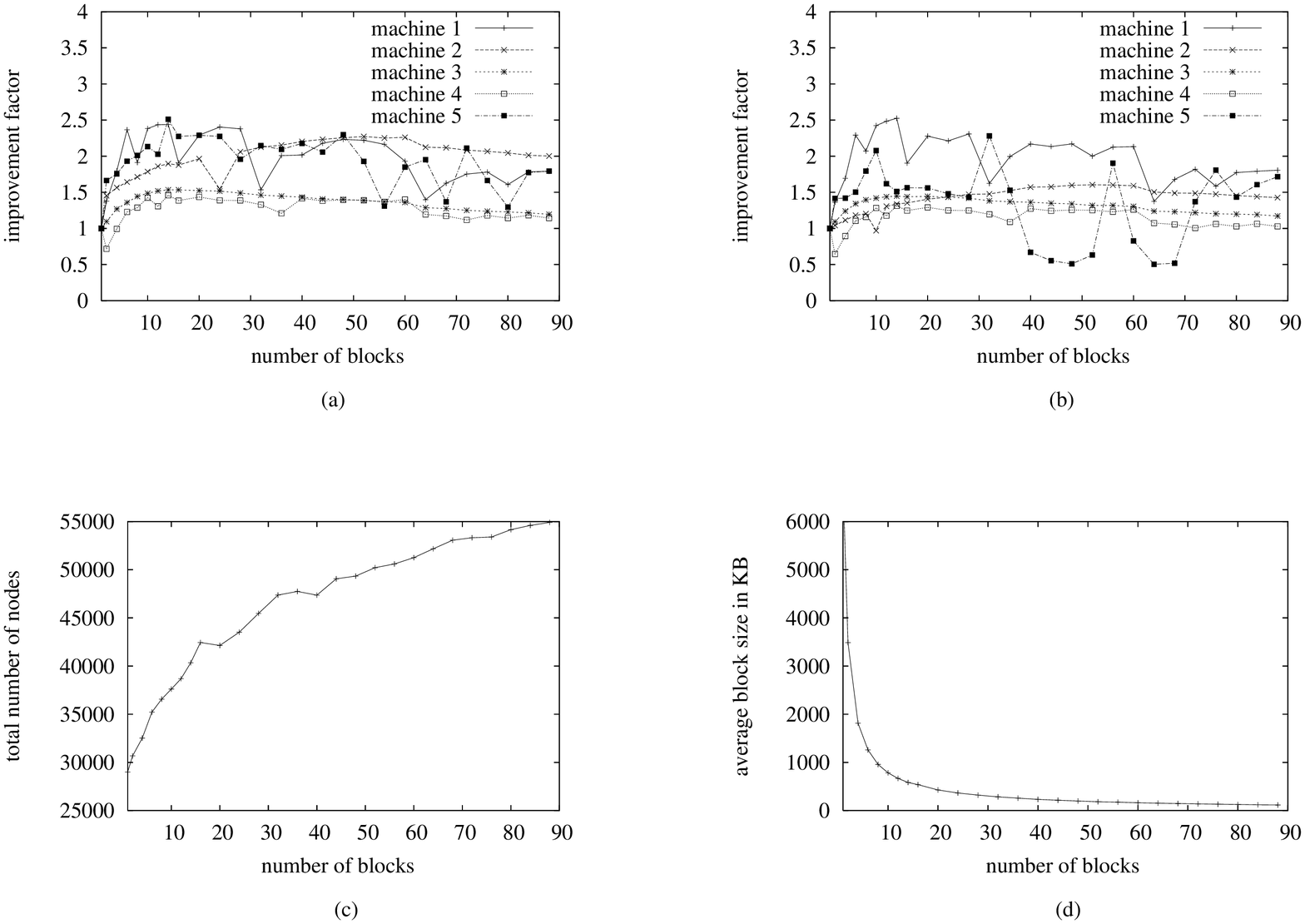}
\caption{Effect of variation of number of blocks on (a) performance
improvement factor of original mesh, (b) performance improvement
factor of reordered mesh, (c) total number of nodes and (d) averaged
block size for case 4.} \label{fig}
\end{center}
\end{figure}

\begin{figure}
\begin{center}
\includegraphics[width=15.cm,height=11.cm]{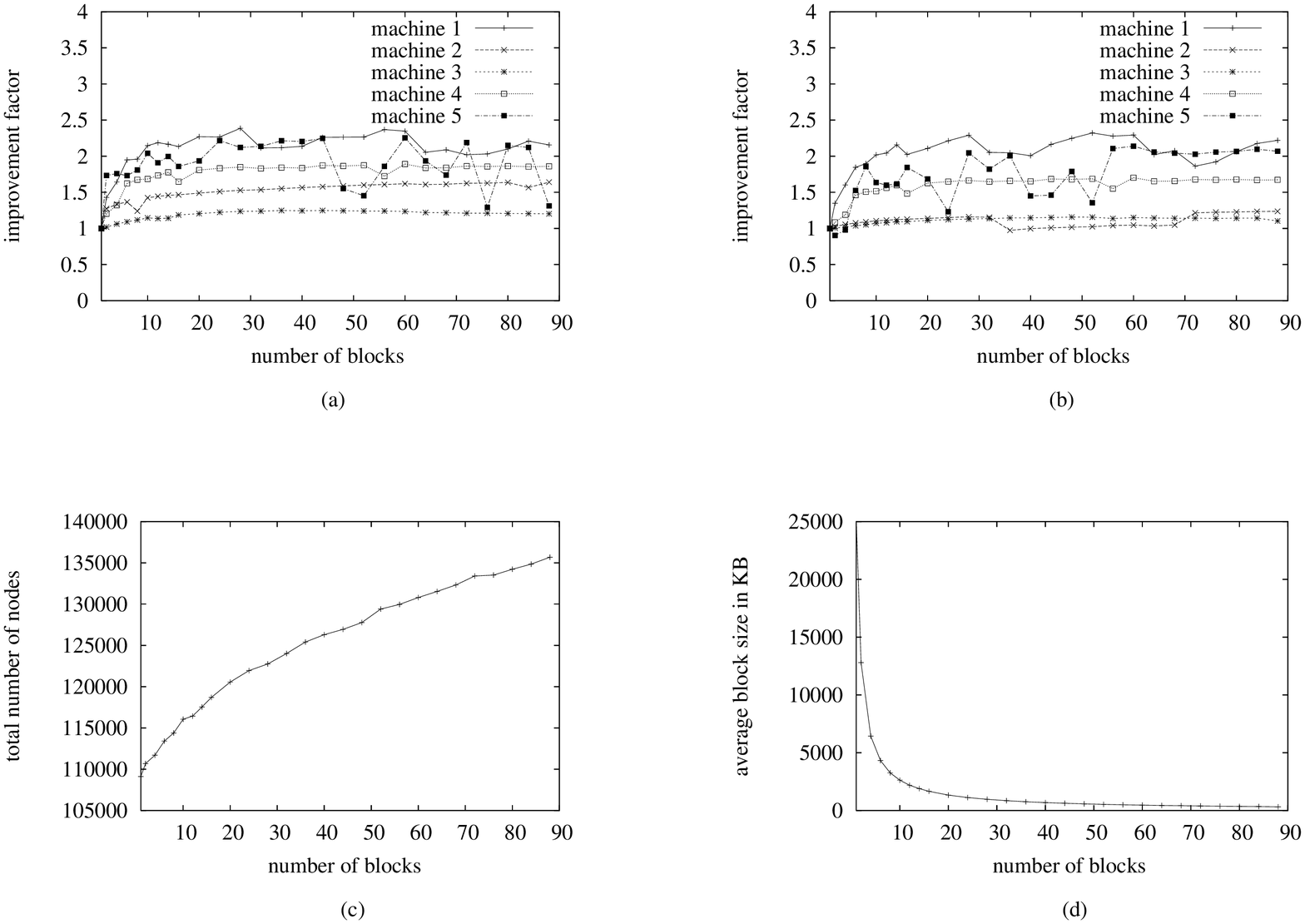}
\caption{Effect of variation of number of blocks on (a) performance
improvement factor of original mesh, (b) performance improvement
factor of reordered mesh, (c) total number of nodes and (d) averaged
block size for case 5.} \label{fig}
\end{center}
\end{figure}

\begin{figure}
\begin{center}
\includegraphics[width=15.cm,height=11.cm]{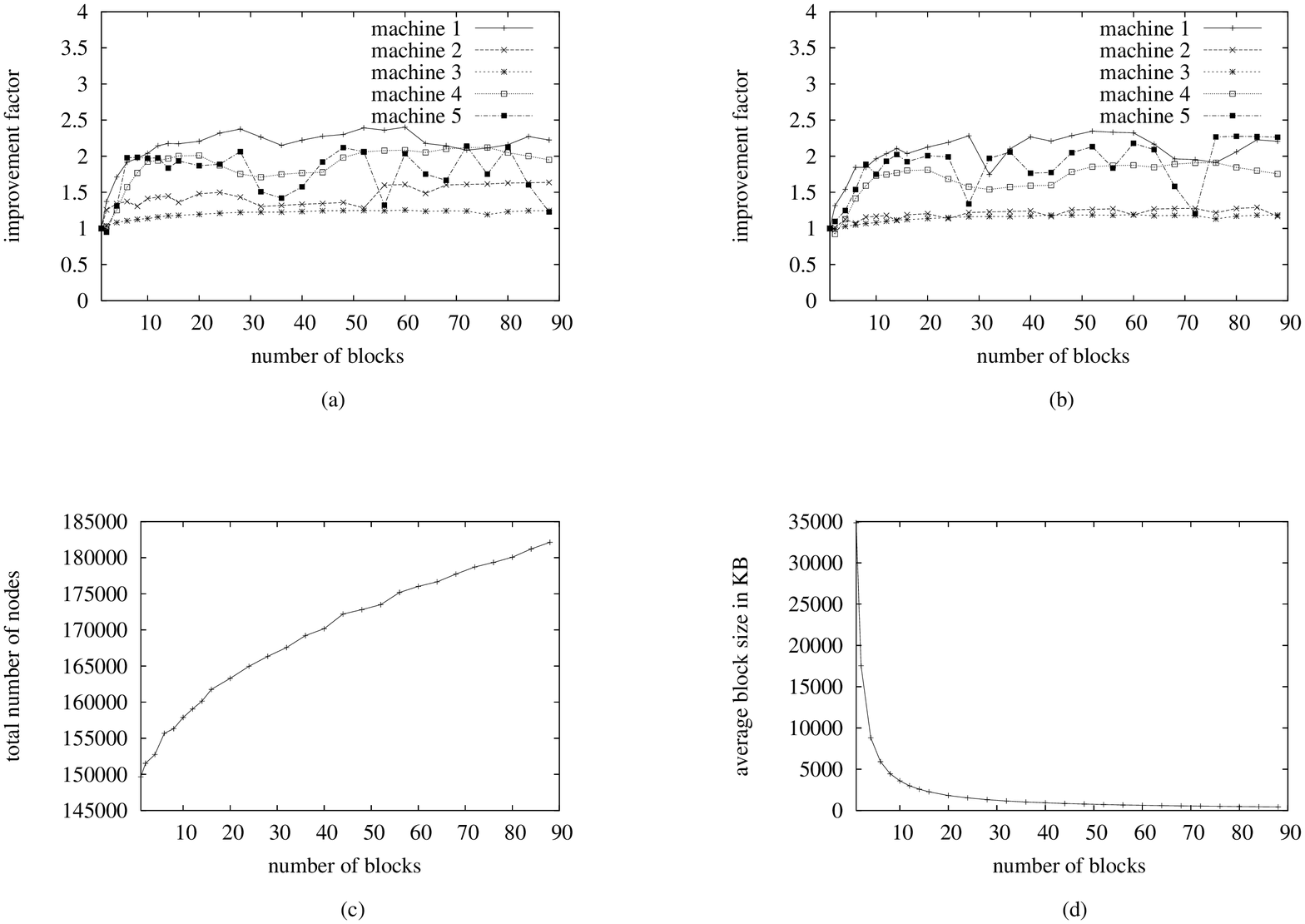}
\caption{Effect of variation of number of blocks on (a) performance
improvement factor of original mesh, (b) performance improvement
factor of reordered mesh, (c) total number of nodes and (d) averaged
block size for case 6.} \label{fig}
\end{center}
\end{figure}

\begin{figure}
\begin{center}
\includegraphics[width=15.cm,height=16.cm]{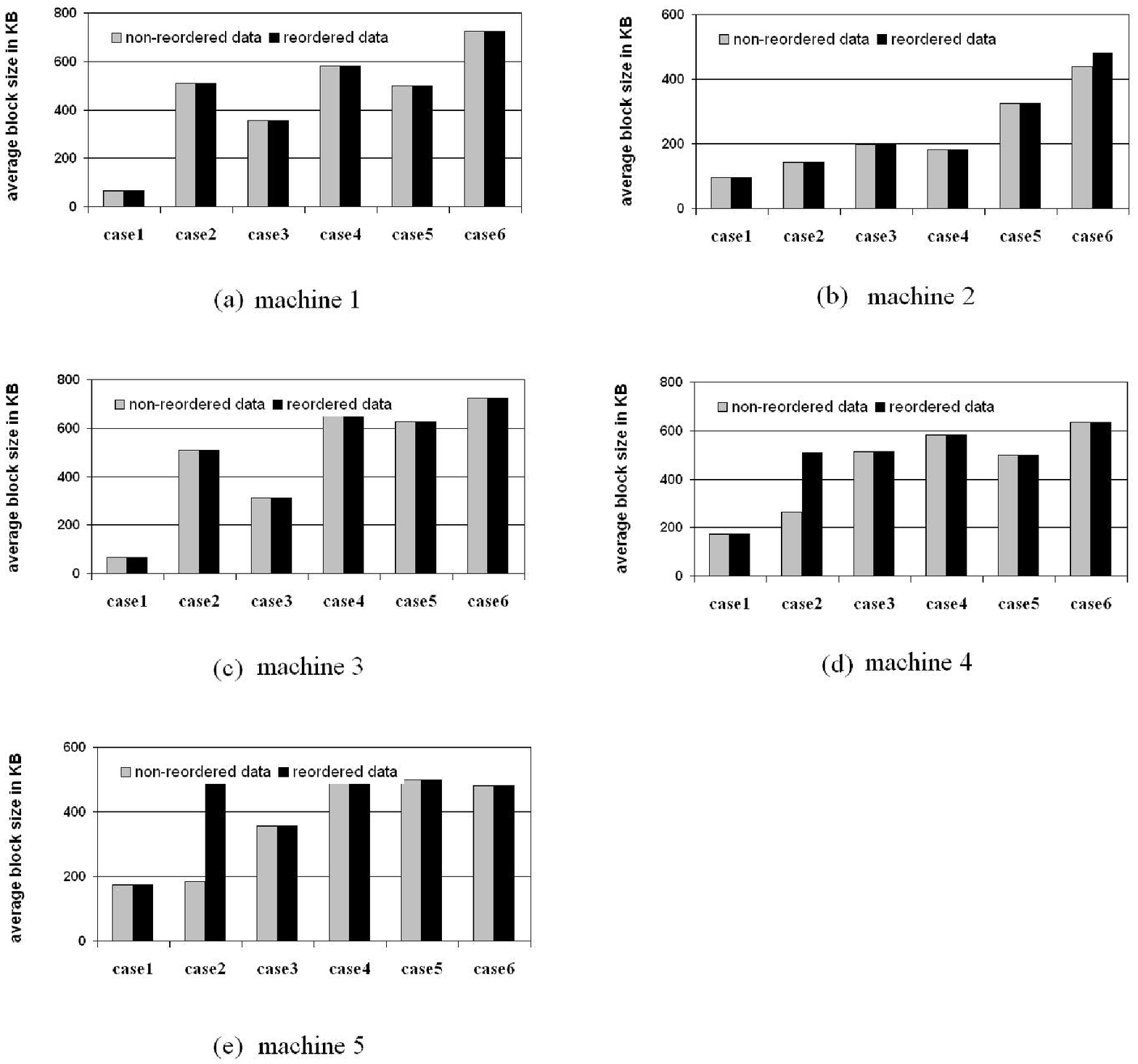}
\caption{The average block size of cases 1-6 at the performance peak
for various machines.} \label{fig}
\end{center}
\end{figure}

\begin{figure}
\begin{center}
\includegraphics[width=13.cm,height=18.cm]{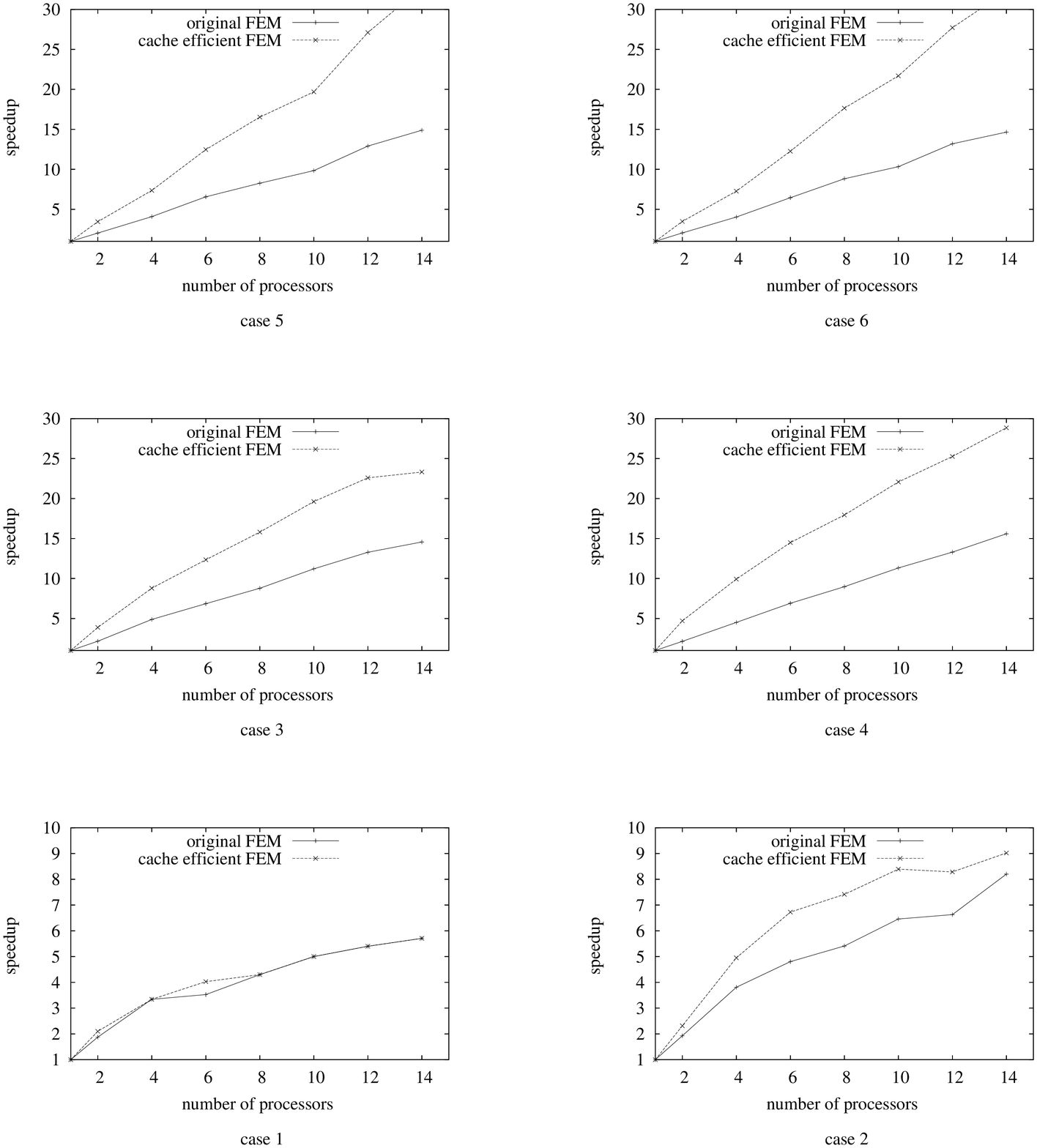}
\caption{Parallel exaction results: speedup curves of traditional
finite element procedure vs. presented cache-efficient ones (CPU
time of traditional method was used as reference CPU time for
calculation of speedup).} \label{fig}
\end{center}
\end{figure}



\end{document}